\documentclass[twocolumn]{aastex62}

\usepackage{subfigure}

\shorttitle{FRB redshift distribution}
\shortauthors{Zhang \& Zhang}

\begin{document}


\title{ \bf The CHIME Fast Radio Burst Population Does Not Track the Star Formation History of the Universe}



\author{Rachel C. Zhang}
\affiliation{Center for Interdisciplinary Exploration \& Research in Astrophysics (CIERA), Evanston, IL 60202, USA}
\affiliation{Department of Physics \& Astronomy, Northwestern University, Evanston, IL 60202, USA}
\author{Bing Zhang }
\affiliation{Department of Physics and Astronomy, University of Nevada, Las Vegas, Las Vegas, NV 89154, USA}

\begin{abstract}
The redshift distribution of fast radio bursts (FRBs) is not well constrained. The association of the Galactic FRB 200428 with the young magnetar SGR 1935+2154 raises  the working hypothesis that FRB sources track the star formation history of the universe. The discovery of FRB 20200120E in association with a globular cluster in the nearby galaxy M81, however, casts doubts on such an assumption. We apply the Monte Carlo method developed in a previous work to test different FRB redshift distribution models against the recently released first CHIME FRB catalog in terms of their distributions in specific fluence, external dispersion measure ($\rm DM_E$), and inferred isotropic energy. Our results clearly rule out the hypothesis that all FRBs track the star formation history of the universe. The hypothesis that all FRBs track the accumulated stars throughout history describes the data better but still cannot meet both the $\rm DM_E$ and the energy criteria. The data seem to be better modeled with either a redshift distribution model invoking a significant delay with respect to star formation or a hybrid model invoking both a dominant delayed population and a subdominant star formation population. We discuss the implications of this finding for FRB source models.
\end{abstract}


\section{Introduction} \label{sec:intro}

The engines that power cosmological fast radio bursts (FRBs) are not well identified. Different types of engines may follow different redshift distributions (e.g. Fig.1 of \cite{zhangrc21}), so constraints on the FRB redshift distribution would offer clues to the origins of FRBs. The discovery of the Galactic FRB 200428 in association with an X-ray burst from the magnetar SGR 1935+2154 \citep{CHIME-SGR,STARE2-SGR,HXMT-SGR,Integral-SGR,Konus-SGR,AGILE-SGR} suggests that at least some FRBs originate from young magnetars produced from deaths of massive stars. As a result, the working hypothesis that the majority of FRB sources follow the star formation history of the universe has been widely adopted in the community. Studies of the host galaxy properties and FRB position offsets from the hosts suggest that the FRB population is generally consistent with such a hypothesis \citep{lizhang20,bhandari20,heintz20,bochenek21,mannings21,fong21}, even though the alternative hypothesis that FRBs follow a stellar population with a significant delay with respect to star formation (e.g. the distribution that track binary neutron star mergers) is not ruled out \citep{lizhang20}. Recently, a repeating source FRB 2020120E was discovered to be located in a globular cluster of the nearby galaxy M81 \citep{bhardwaj21,kirsten21,nimmo21}, suggesting that at least some FRBs are associated with old stellar populations. It is desirable to know whether such a delayed population makes up a significant fraction of the entire FRB population. 

So far, the FRB redshift distribution is poorly constrained. Only more than a dozen FRBs have direct redshift measurements \citep{tendulkar17,bannister19,ravi19,prochaska19,marcote20,macquart20,bhandari21}. These bursts were detected with a variety of radio telescopes that have very different instrumental selection effects, so this limited sample cannot give a reliable constraint on FRB redshift distribution. The dispersion measure (DM) of an FRB can be a rough proxy of its redshift, as is verified by the observational confirmation \citep{macquart20} of the ${\rm DM}-z$ relation long suggested theoretically \citep{ioka03,inoue04,deng14}. In principle, based on the observed fluence distribution, DM distribution, and inferred energy distribution of FRBs, one can constrain the redshift distribution of FRBs \citep{zhangrc21,james21}. However, the situation is so far inconclusive due to the small FRB samples and complicated observational selection effects. In particular, the small Parkes and ASKAP FRB samples are not inconsistent with either a model tracking the star formation history of the universe \citep{zhangrc21,james21} or models invoking a significant delay with respect to star formation (e.g. those related to binary neutron star mergers) \citep{zhangrc21}.

Recently, the CHIME/FRB Collaboration published their first FRB catalog \citep{chime-1st-catalog} reporting 536 FRBs including 62 bursts detected from 18 repeating sources. This uniform large sample provides an ideal resource to constrain the FRB redshift distribution. \cite{chawla21} performed a population study of the CHIME catalog. They focused on DM and scattering distributions and constrained the properties of the circumgalactic medium. They only assumed that the FRB rate evolves with redshift and tracks the star formation history of the universe without testing a range of redshift distribution models. 

In this Letter, following the Monte Carlo method described in our earlier work \citep{zhangrc21}, we systematically investigate the consistency of various redshift distribution models with the CHIME FRB catalog. We rule out the hypothesis that the entire FRB population tracks the star formation history of the universe with high significance. We also reject the hypothesis that the FRB population tracks the accumulated stars throughout the history of the universe, despite appearing to match the data better. The data seem to instead approach a model that either requires the FRB sources to have a significant delay with respect to star formation or a hybrid model that includes both a dominant delayed population and a subdominant star formation population. Our method is reviewed in Section \ref{sec:method}. The results are presented in Section \ref{sec:results}, and conclusions are drawn in Section \ref{sec:conclusions} with some discussion. 

\section{The Method}\label{sec:method}

The details of our Monte Carlo method have been described in \cite{zhangrc21}. Here, we only outline the key ingredients of the method. Basically, for a uniform FRB sample detected by the same telescope (as is the case of the CHIME sample), the fluence, distance (and hence, $\rm DM_E$), and energy distributions of the observed FRB population depend on three factors: { (1) the intrinsic redshift distribution, (2) the intrinsic FRB isotropic energy (or luminosity) distribution, and (3) the telescope's sensitivity threshold and instrumental selection effects near the threshold. }

The first factor is the focus of investigation of this paper. Interestingly, the other two factors are largely decoupled from the first factor and from each other, which allows us to treat them independently. 
The second factor (energy distribution) has been well constrained from observations. Independent studies regardless of the assumed redshift distribution \citep[e.g.][]{luo18,luo20,lu19b,lu20,zhangrc21} have reached the consistent conclusion that the energy distribution of the entire FRB population is roughly a power law $dN/dE \propto E^{-\alpha}$ covering at least 8 orders of magnitude, with the index $\alpha \sim (1.8-2.0)$. There might be a high-energy exponential cutoff \citep{luo20,lu20} but the cutoff energy $E_c$ is not well constrained \citep{zhangrc21}\footnote{Even though the energy distribution was constrained from the observations using the telescopes (e.g. Parkes, ASKAP in $\sim$ GHz) with observing bands higher than that of CHIME (400 to 800 MHz), it is reasonable to assume that the {\em shape} of the $E$ distribution in the CHIME band is similar to that constrained in the GHz band. Throughout the paper, our $E$ is defined as the isotropic energy as observed in the CHIME frequency band.}.  In principle, there could be a redshift evolution of FRB energy distribution. However, current data do not require such an evolution. Furthermore, the mechanism of FRB sources (e.g. magnetars) is likely related to their own physical properties rather than redshift. As a result, we have assumed a universal energy function for all FRBs throughout the universe.  The third factor (instrumental effects) is difficult to characterize. The CHIME catalog data show that the telescope specific fluence  cutoff is about $0.3 \ {\rm Jy \ ms}$, or $\log {\cal F}_{\rm min} \simeq -0.5$ (see panel (a) of Figs.\ref{fig:SFH}-\ref{fig:hybrid}). However, due to many instrumental or human-related uncertainties of CHIME observations (e.g. unknown positions of most bursts that introduce large errors in the estimated fluences, nonuniform CHIME sensitivity on the sky due to large gaps between beams, and DM dependence and scattering time dependence of the detection efficiency \citealt{chime-1st-catalog}), there is a ``gray zone'' in the ${\cal F}$ distribution within which the CHIME telescope has not reached full sensitivity to all sources. These effects are independent of the redshift distribution and may be corrected using an empirical model (see below)\footnote{Effectively, this is to assume that the efficiency for CHIME to detect FRBs is essentially independent of the $\rm DM$ value of the FRB, especially around the $\rm DM_E$ distribution peak region $\sim (250 - 500) \ {\rm pc \ cm^{-2}}$. This is consistent with the analysis by the CHIME team, who stated that ``the selection effects in DM are modest'' and that ``we appear to be detecting the full range of DMs represented in the population detectable at CHIME/FRB's sensitivity'' \citep{chime-1st-catalog}.}.

By adopting an intrinsic energy ($E$) distribution and a redshift ($z$) distribution, one can simulate a large number of mock FRBs. The specific fluence of each mock burst can be calculated based on its assigned $E$ and $z$ values. After screening them using a telescope sensitivity model, one can finally obtain a mock ``observed'' sample of FRBs. Based on the ${\rm DM}-z$ relation \citep{deng14,zhang18b,pol19,cordes21}
\begin{equation}
    {\rm DM_{IGM}}(z) = \frac{3c H_0 \Omega_b f_{\rm IGM}}{8\pi G m_p} \int_0^z \frac{(7/8)(1+z) dz}{\sqrt{\Omega_m(1+z)^3+\Omega_\lambda}}
\end{equation}
(where $H_0$, $\Omega_b$, $\Omega_m$, $\Omega_\lambda$ are cosmological parameters whose values are adopted from the latest Planck results \citep{planck}, $G$ is the gravitational constant, $m_p$ is proton mass, and $f_{\rm IGM}$ is the fraction of baryons in the IGM, which is adopted as 0.84), one can estimate the IGM portion of the dispersion measure ($\rm DM_{IGM}$) of each mock FRB. Adopting a model for the host galaxy dispersion measure ($\rm DM_{host}$), one can finally simulate the excess DM distribution of the mock sample, which can be compared with the $\rm DM_E = DM_{IGM}+DM_{host}/(1+z)$ data directly retrievable from the CHIME catalog \citep{chime-1st-catalog}\footnote{The $\rm DM_E$ can be obtained by subtracting the Milky Way contribution $\rm DM_{\rm MW}$ and the Milky Way halo contribution $\rm DM_{halo}$ from the measured DM. The former is derived from the MW electron density models NE2001 \citep{cordes02} or YMW16 \citep{yao17} and we adopt the NE2001 model throughout the paper. For the latter, we assume $\rm DM_{halo} \sim 30 \ {\rm pc \ cm^{-3}}$ for all FRBs \citep[e.g.][]{dolag15,prochaska19b}.}. 

In order to test a certain $z$-distribution model, we make use of three observational criteria (see Figs.\ref{fig:SFH}-\ref{fig:hybrid}): (1) the specific fluence ($\log {\cal F}_\nu$) distribution, (2) the isotropic energy ($\log E$) distribution, and (3) the excess dispersion measure ($\rm DM_E$) distribution. The specific fluence is a convolution of the energy and redshift distributions and is insensitive to either distribution.\footnote{The insensitivity of fluence distribution on redshift distribution was clearly revealed by the nearly identical fluence distributions in the Fermi GBM catalog \citep{paciesas12} for both long and short gamma-ray bursts, which have very different redshift distributions.} This is because the fluence distribution (also called $\log N - \log {\cal F}_\nu$ distribution) should follow a simple $N \propto {\cal F}_\nu^{-3/2}$ distribution regardless of the energy function if the sources are uniformly distributed in a Euclidean space.\footnote{This is because for a given specific burst energy and a uniform number density in space, the specific fluence is proportional to $r^{-2}$ and the total number is proportional to $r^3$, and because such scaling relations remain the same for different burst energies \citep[e.g.][]{zhang18}.} The non-Euclidean geometry of cosmological models may break the simple scaling, but only in the low fluence regime. However, in this low fluence regime, the instrumental selection effects become so important that any redshift-distribution-related features are removed. As a result, the criterion 1 is most easily satisfied among all three criteria for all models. For every pair of $z$ and $E$ distribution models, it is possible to find an empirical instrumental selection effect model to satisfy the $\log {\cal F}_\nu$ criterion. However, many models that satisfy the $\log {\cal F}_\nu$ distribution criterion could fail the $\log E$ and $\rm DM_E$ criteria. We therefore come up with the following strategy: for each $z$-distribution model, we adjust the $E$ distribution model and the empirical sensitivity model to make the ${\cal F}_\nu$ distribution of the mock ``observed'' sample not be rejected by the Kolmogorov--Smirnov (K-S) test (all the K-S test statistics in this paper are reported with 95\% confidence) against the observed ${\cal F}_\nu$ distribution. We then go on to evaluate the $\log E$ and $\rm DM_E$ distribution criteria. The model is ruled out if the same mock FRB sample fails both criteria. 

For instrumental sensitivity threshold modeling in the ``gray zone'' of the lower end of the specific fluence distribution, we adopt $\log {\cal F}_{\rm \nu,th}^{\rm min} = -0.5$ as the minimum threshold specific fluence (as shown by the data). We define a maximum threshold specific fluence, $\log {\cal F}_{\rm \nu,th}^{\rm max}$, whose value is adjusted to match the observation, and define a detection efficiency parameter, $\eta_{\rm det}$, that depends on the ratio ${\cal R} = (\log {\cal F}_{\rm \nu,th} - \log {\cal F}_{\rm \nu,th}^{\rm min}) / (\log {\cal F}_{\rm \nu,th}^{\rm max} - \log {\cal F}_{\rm \nu,th}^{\rm min})$ for fluences between $\log {\cal F}_{\rm \nu,th}^{\rm min}$ and $\log {\cal F}_{\rm \nu,th}^{\rm max}$, such that $\eta_{\rm det} \rightarrow 0$ at $\log {\cal F}_{\rm \nu,th}^{\rm min}$ and $\eta_{\rm det} \rightarrow 1$ at $\log {\cal F}_{\rm \nu,th}^{\rm max}$. We model the dependence as $\eta_{\rm det} = {\cal R}^n$ and find $n=3$ can model the $\log {\cal F}_\nu$ distribution and adopt this empirical function form and the typical value in our modeling. 

Strictly speaking, a more rigorous but cumbersome method would be to vary all the free parameters (for the intrinsic energy distribution model and the empirical sensitivity model) for each redshift distribution model to compare against all three observational criteria. However, a salient feature is that both the fluence distribution and the instrumental effects near the fluence sensitivity threshold essentially do not depend on the redshift distribution model of FRBs. This allows us to adopt the abovementioned simpler approach to test all the $z$-distribution models. Nonetheless, for the star formation history model, which is the main model of interest, we also apply the more rigorous method to confirm the validity of the simpler method. {We test the range $(1.8, 2.0)$ inclusive with step size 0.1 for energy power-law index $\alpha, (41.5, 43.5)$ inclusive with step size 1 for $\log E_c, (0.65, 0.85)$ inclusive with step size 0.1 for ${\cal F}_{\rm \nu,th}^{\rm max}$, and (1, 3) inclusive with step size 1 for the index $n$ in the instrumental empirical function. We find a total of 15 models that are not rejected by the 1D K-S test for the ${\cal F}_\nu$ distribution. All 15 of those models, however, are rejected by the 1D K-S test for both the energy and $\rm DM_E$ distributions. Thus, as expected, even with the best sets of parameters, the model fails to reproduce the data, so our main conclusion of the paper, i.e. FRBs do not track the star formation history of the universe, is robust.} Since our purpose is not to find the best parameters to meet the data, and since there are more parameters in the delayed and hybrid models, we do not perform a multi-dimensional parameter search for those models.

In our simulations, we use the central value of $\rm DM_{host}$,  $107 \ {\rm pc \ cm^{-3}}$, as constrained from the data \citep[e.g.][]{lizx20}. In principle, both the $\rm DM_{host}$ and $\rm DM_{IGM}$ values for individual FRBs should have a distribution around the central values. However, since we are simulating a large sample of mock FRBs, the resulting {\em distributions} of various parameters by adopting central values should be similar to the more realistic case involving these distributions. {In particular, our treatment of $\rm DM_{host}$ is valid if its true distribution is normal. The true $\rm DM_{host}$ distribution is, however, difficult to constrain from current observations and may have a more complicated form. For example, some FRBs have very large $\rm DM_{host}$ values \citep{niu21}. Nonetheless, a larger $\rm DM_{host}$ would imply a smaller $\rm DM_{IGM}$ and a smaller $z$.  If a larger $\rm DM_{host}$ value than assumed is prevailing among FRBs, the true FRB $z$ distribution would have a lower peak than what is inferred in the following discussion. This only strengthens the conclusion drawn in Section \ref{sec:results}.}

\section{Results}\label{sec:results}

We investigate four families of $z$-distribution models in detail: 
\begin{enumerate}
    \item Star formation rate history (SFH) model: FRB sources follow the star formation history of the universe;
    \item Accumulated model: FRB sources follow the number of total stars in the universe;
    \item Delayed model: FRB sources follow stellar populations that have a significant delay with respect to star formation;
    \item Hybrid model: a fraction of FRB sources follows star formation and another fraction follows a delayed population.
\end{enumerate}

\begin{figure*}
    \centering
    \includegraphics[width=\textwidth]{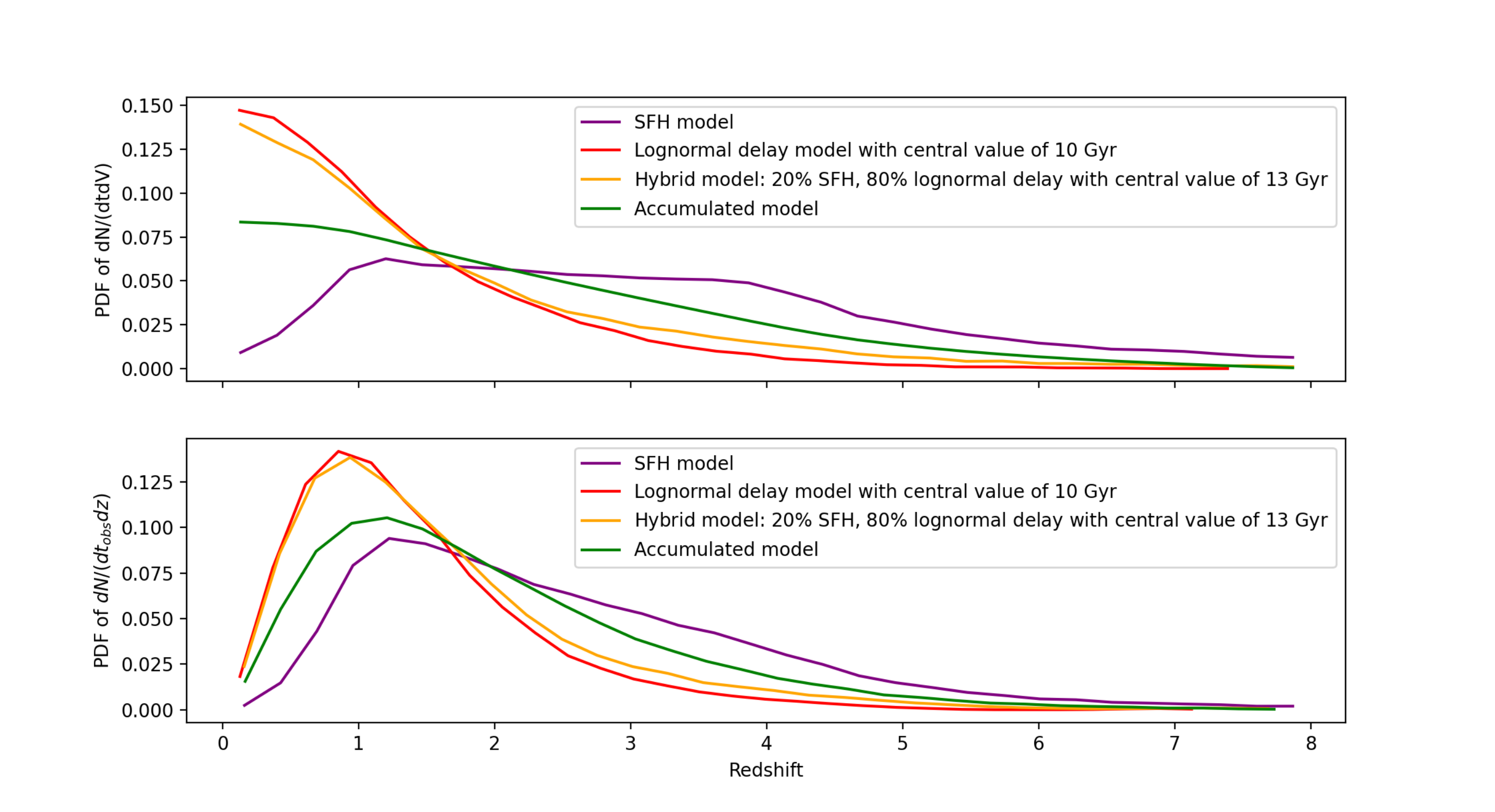}
    \caption{Probability distribution functions (PDFs) for the SFH model, lognormal delay model with a central value of 10 Gyr, hybrid model with a 20\% contribution from the SFH model and 80\% from a lognormal delay model with a central value of 13 Gyr, and accumulated SFH model. The top panel shows the PDF of the intrinsic FRB event-rate density  $dN/(dtdV)$, while the bottom panel shows the PDF of the observed FRB event-rate redshift distribution $dN/(dt_{\rm obs}dz)$.}
    \label{fig:models}
\end{figure*}

For each model, we first build a model of their redshift-dependent event-rate density $dN/(dV dt)$. We then convert it to a redshift rate distribution using \citep[e.g.][]{sun15}
\begin{equation}
    \frac{dN}{d t_{\rm obs} dz} = \frac{1}{1+z} \frac{dN}{dt dV} \frac{dV}{dz},
\label{eq:dN/dtdz}
\end{equation}
where 
\begin{equation}
    \frac{dV}{dz} = \frac{c}{H_0}\frac{4\pi D_{\rm L}^2}{(1+z)^2 \sqrt{\Omega_m(1+z)^3}+\Omega_\lambda},
\end{equation}
and $D_{\rm L}$ is the luminosity distance at $z$. Since we only care about the $z$ distribution, we use the normalized probability distribution functions (PDFs) of $dN/(dt dV)$ and $dN/(dt_{\rm obs} dz)$ for all of the models, which are presented in the upper and lower panels, respectively, of Figure \ref{fig:models}. For the data, we take all nonrepeaters and only the first detected burst for each repeating FRB from the CHIME catalog. 

{For each model, we provide example parameters we use to get the respective fluence distributions not rejected by the K-S test in Table \ref{table1}. We note that the SFH model requires a higher $E_c$ than the other models because the SFH model predicts on average higher redshift FRBs, which requires higher energies to get the same fluence distribution. }

\begin{table*}
	\centering
	\begin{tabular}{c|c|c|c|c}
	\hline
	 &$\alpha$& $\log E_c$ & $\log {\cal F}_{\rm \nu,th}^{\rm max}$ & $n$ \\
	\hline
	SFH & 1.9 & 41.5 & 0.85 & 3 \\
	\hline
	Accumulated & 1.9 & 41 & 0.85 & 3 \\
	\hline
	Delayed & 1.9 & 41 & 0.85 & 3  \\
	\hline 
	Hybrid & 1.9 & 41 & 0.85 & 3  \\
	\hline
	\end{tabular}
	\label{table1}
	\caption{Example parameters for the four models that create fluence distributions that are not rejected by the K-S test.} 
\end{table*}

\subsection{Star formation history (SFH) model}

This is the most well-motivated model. Galactic magnetars are usually believed to be young neutron stars born from supernova explosions, some of which are found to be associated with young star clusters or supernova remnants \citep{kaspi17}. If magnetars are the sources of most FRBs in the universe \citep[e.g.][]{lizhang20,bhandari20,heintz20,bochenek21}, one would then naturally expect that FRBs follow the star formation history of the universe \citep{zhangrc21,james21,chime-1st-catalog}.

To test this model, we use the analytical three-segment empirical model of \cite{yuksel08}. This model is consistent with the widely used two-segment empirical model of \cite{madau14} but more precisely catches the SFH at high redshifts mapped by long gamma-ray burst observations. We assume that $dN/(dt dV)$ of FRBs is proportional to the volumetric star formation rate and derive its redshift distribution $dN/(dt_{\rm obs} dV)$ (purple curves in Fig.\ref{fig:models}). 

\begin{figure*}
    \centering
    \subfigure[]{\includegraphics[width=0.39\paperwidth]{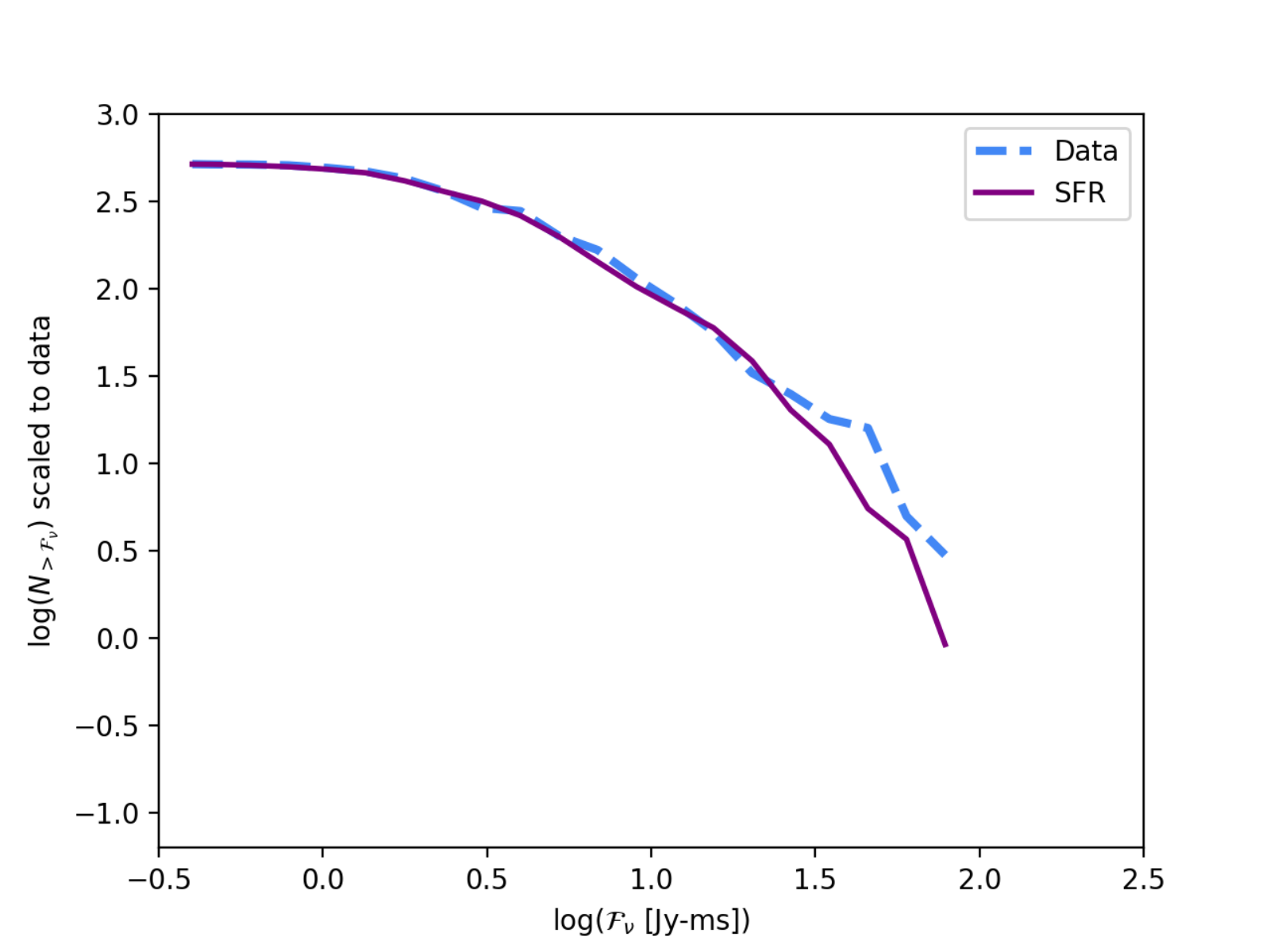}} 
    \subfigure[]{\includegraphics[width=0.39\paperwidth]{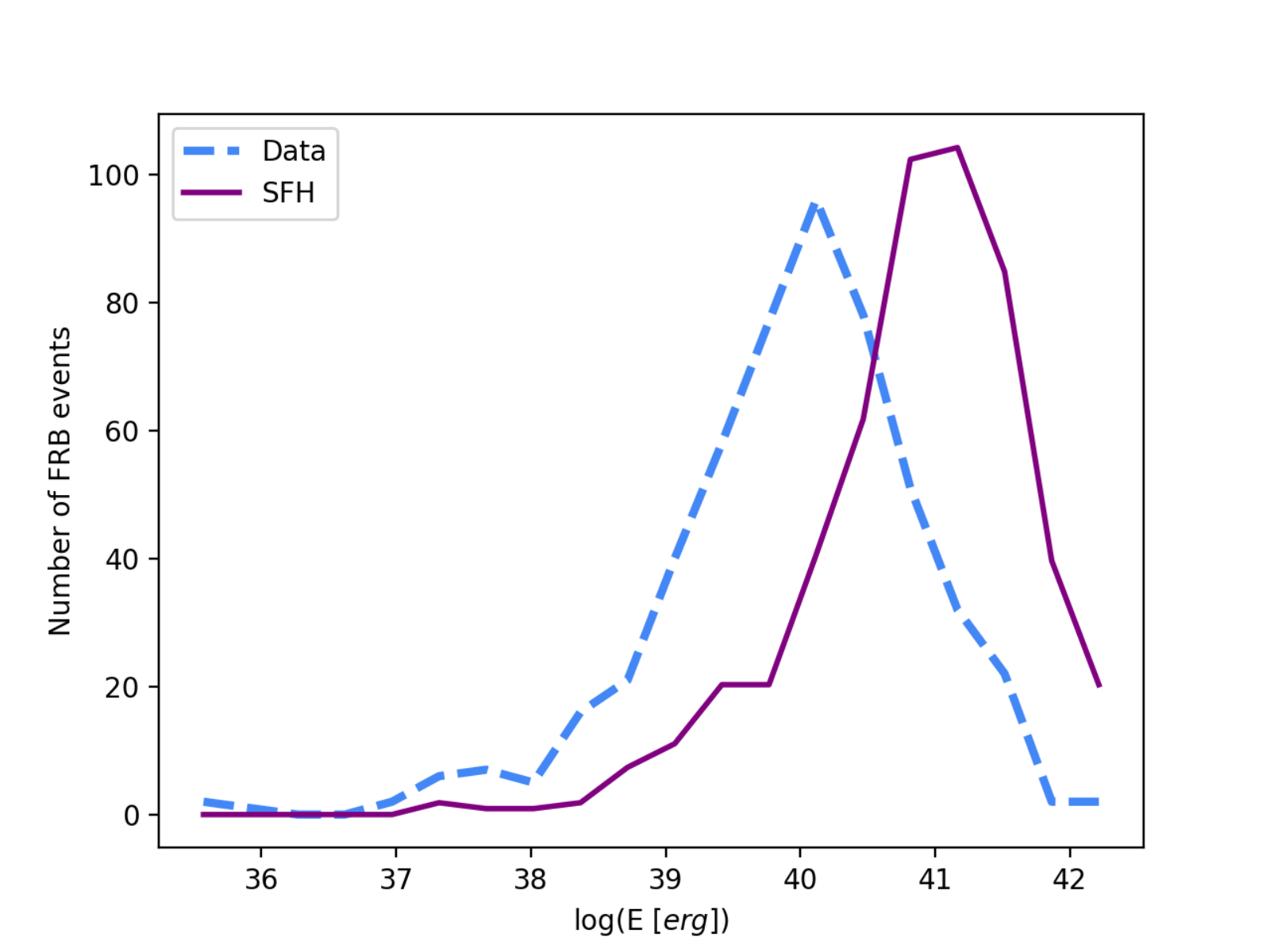}} 
    \newline
    \subfigure[]{\includegraphics[width=0.39\paperwidth]{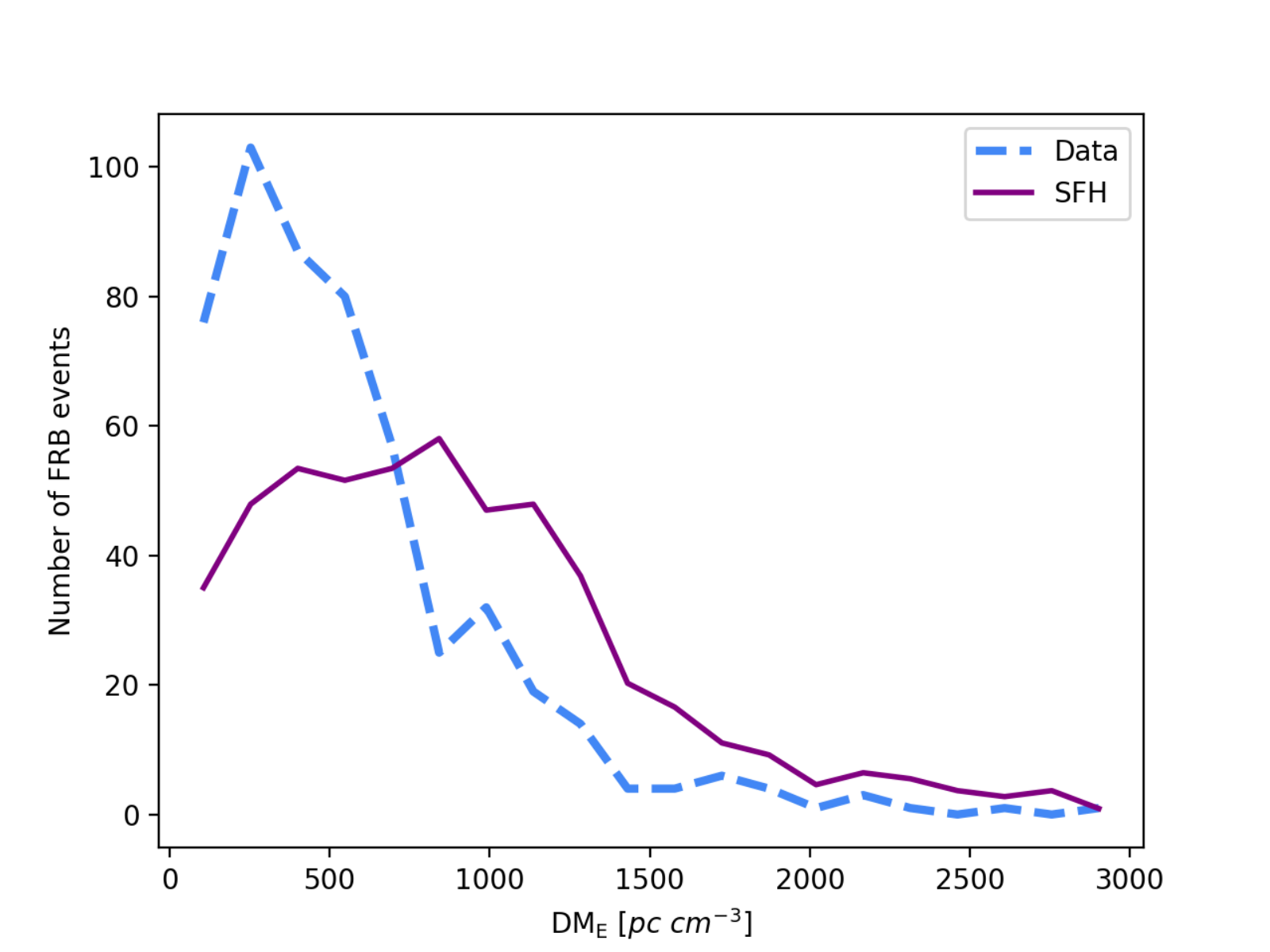}}
    \subfigure[]{\includegraphics[width=0.39\paperwidth]{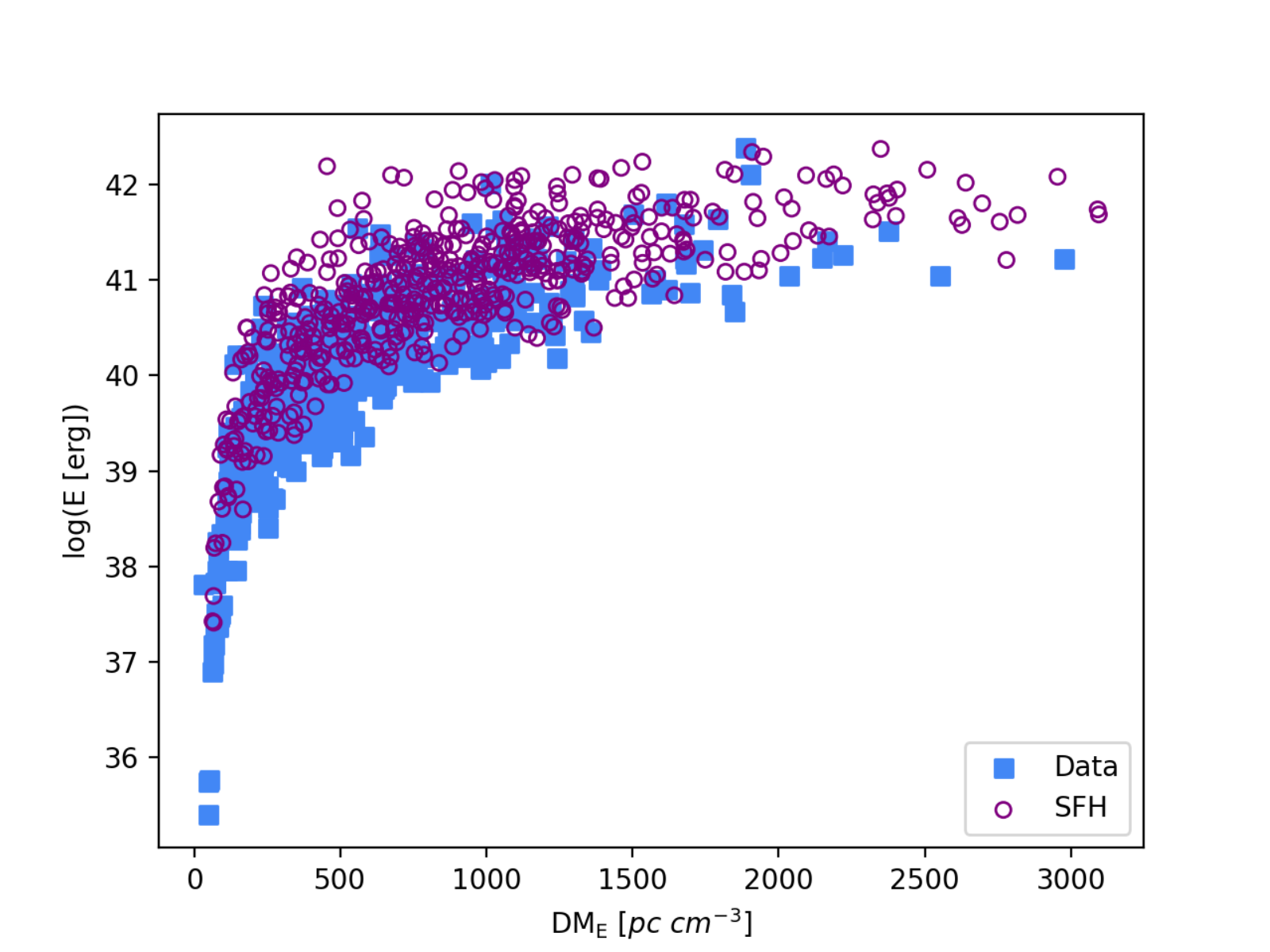}}
    \caption{A test of the SFH model against the three observational criteria. The full FRB sample is tested against the model predictions. (a) The $\log{N}_{>{\cal F}_\nu} - \log{\cal F}_\nu$ test, (b) the $\log E$ distribution test, (c) the $\rm DM_E$ distribution test, and (d) the 2D $\rm DM_E - \log E$ distribution for illustrative purpose (not for testing). For panels (a), (b), and (c), the SFH simulations are scaled to the full data. For the energy distribution model, we adopt $\alpha = 1.9$ and $\log E_c = 41.5$.}
\label{fig:SFH}
\end{figure*}

Even though this model was found consistent with the smaller Parkes and ASKAP FRB samples \citep{zhangrc21,james21}, Figure \ref{fig:SFH} shows that it fails to account for the CHIME data. The model overpredicts the number of FRBs at relatively high redshifts, and hence, high $\rm DM_E$ (Fig.\ref{fig:SFH}c)\footnote{Note that even though the fluence and energy distributions are plotted in the logarithmic scale, the $\rm DM_E$ distribution, which is a proxy of the $z$ distribution, is plotted in the linear scale. This is because unlike ${\cal F}_{\nu}$ and $E$ which have power-law distributions, none of the physical parameter scales with $z$ as a power law. A more relevant parameter to delineate power-law behavior in cosmology is $(1+z)$, which is inversely proportional to the scale size of the universe. This parameter only changes by a factor of 2 from $z=0$ to $z=1$. For the redshift range we are interested in, it is more reasonable to present the $\rm DM_E$ distribution in the linear space.}, which requires an $E$ distribution that has a higher peak than observed (Fig.\ref{fig:SFH}b). The K-S tests for these two criteria show very clearly that the star formation history model is rejected to describe the data. In Fig.\ref{fig:SFH}d, we show the two-dimensional distribution of the data and simulated mock FRBs in the ${\rm DM_E} - \log E$ space. A deficit of low $\rm DM_E$, low $\log E$ FRBs is clearly seen. 

Since the SFH model is the most speculated and discussed model, we also perform two additional tests. First, even though both the 1D K-S tests for $\rm DM_E$ and $E$ criteria have rejected the model, we still perform a 2D K-S test for fluence and $\rm DM_E$ to see whether the model has any chance of survival. We find that the null hypothesis that the data and star formation rate history simulation are from the same inherent distribution is rejected. Second, we also perform the more rigorous method as discussed in Section \ref{sec:method} to explore the compliance of the model with the data from multidimensional parameter space. Our results show that except for a small parameter space where K-S tests can pass the fluence distribution criterion but fail both $\rm DM_E$ and $E$ criteria, in the majority of the parameter space, all three criteria fail. These two additional tests strengthen our conclusion that the hypothesis that the CHIME FRB population track the star formation history of the universe is ruled out by the data with high significance. {The main reason that this model fails to meet the data is that the observed $dN/ d {\rm DM_E}$ distribution drops significantly above $\sim 250 \ {\rm pc \ cm^{-2}}$, which corresponds to $z \sim 0.3$. This is in sharp contrast to the prediction of the SFH model (see Fig. \ref{fig:models}). This demands producing many more FRBs in the nearby universe than what the SFH model predicts, which requires various delayed models as discussed below. }

\subsection{Accumulated model}

\begin{figure*}
    \centering
    \subfigure[]{\includegraphics[width=0.39\paperwidth]{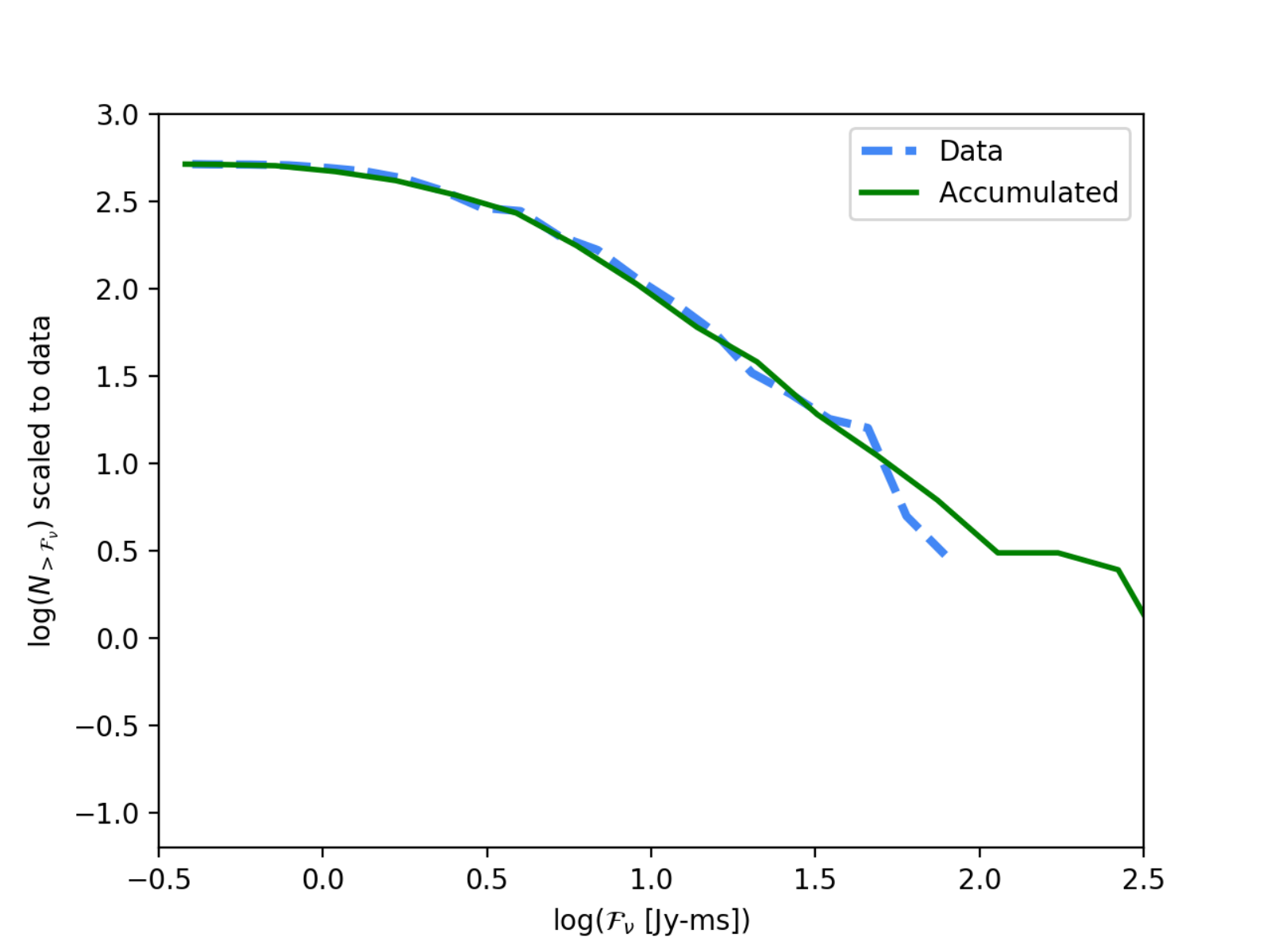}} 
    \subfigure[]{\includegraphics[width=0.39\paperwidth]{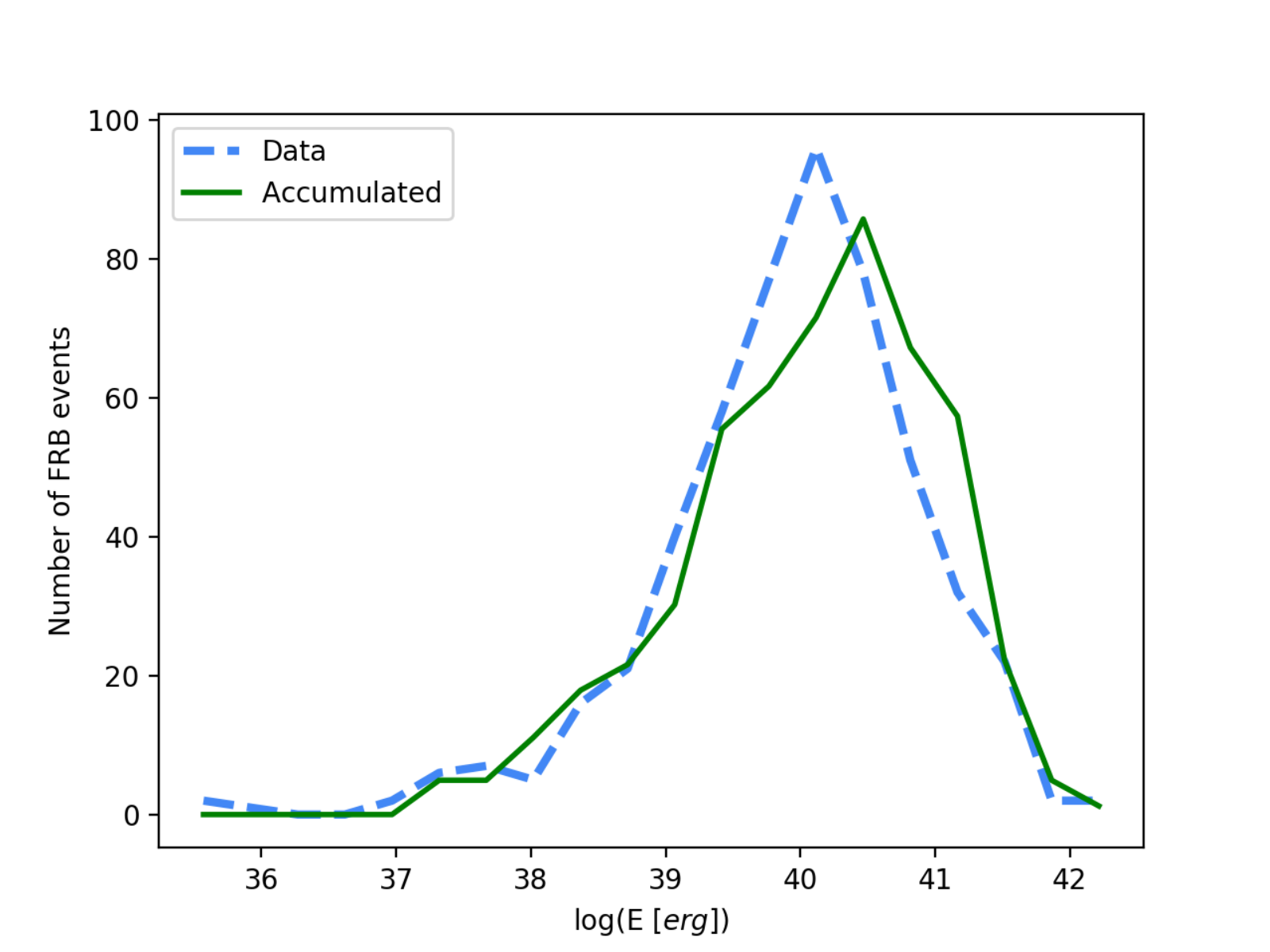}} 
    \newline
    \subfigure[]{\includegraphics[width=0.39\paperwidth]{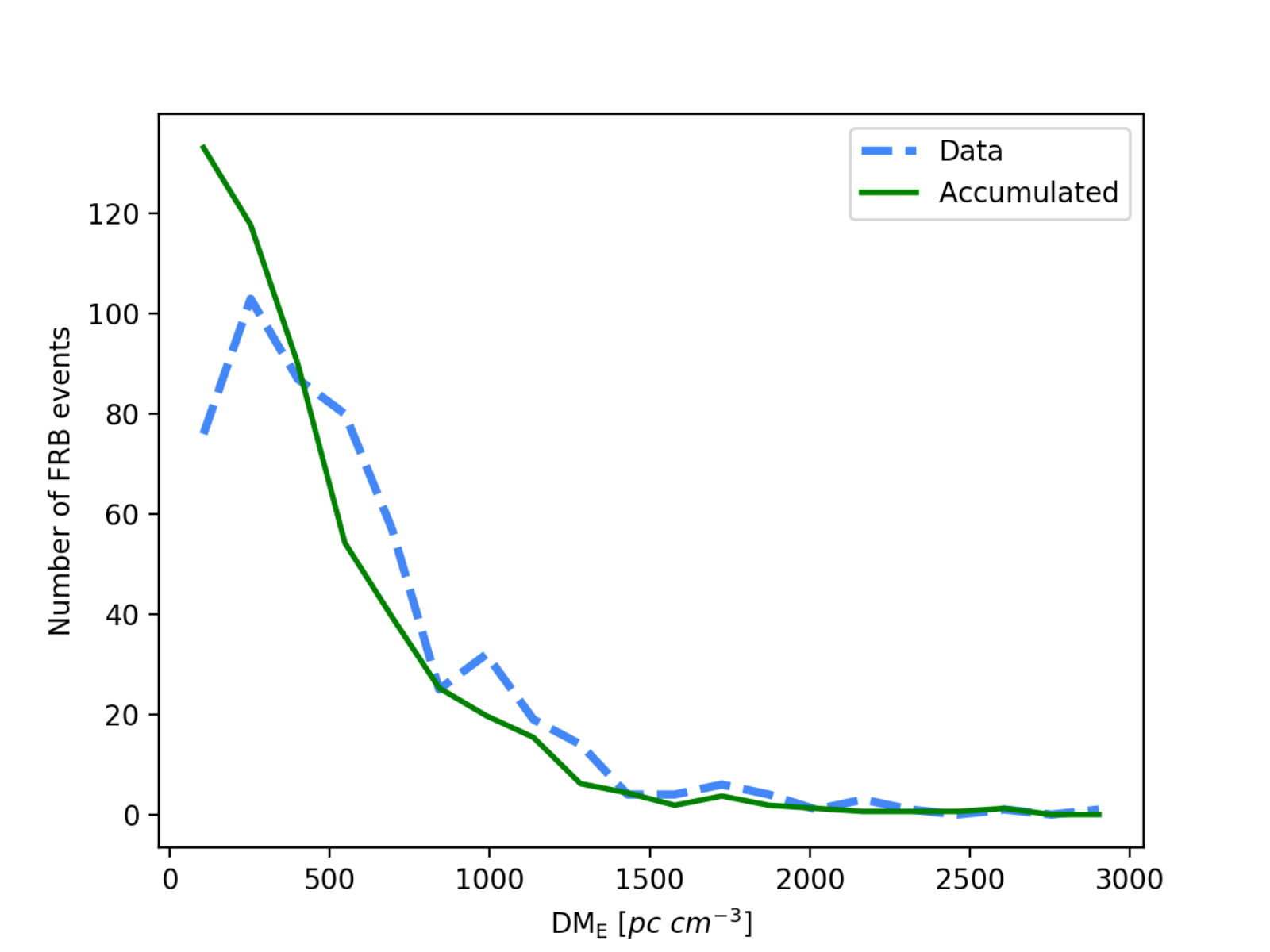}}
    \subfigure[]{\includegraphics[width=0.39\paperwidth]{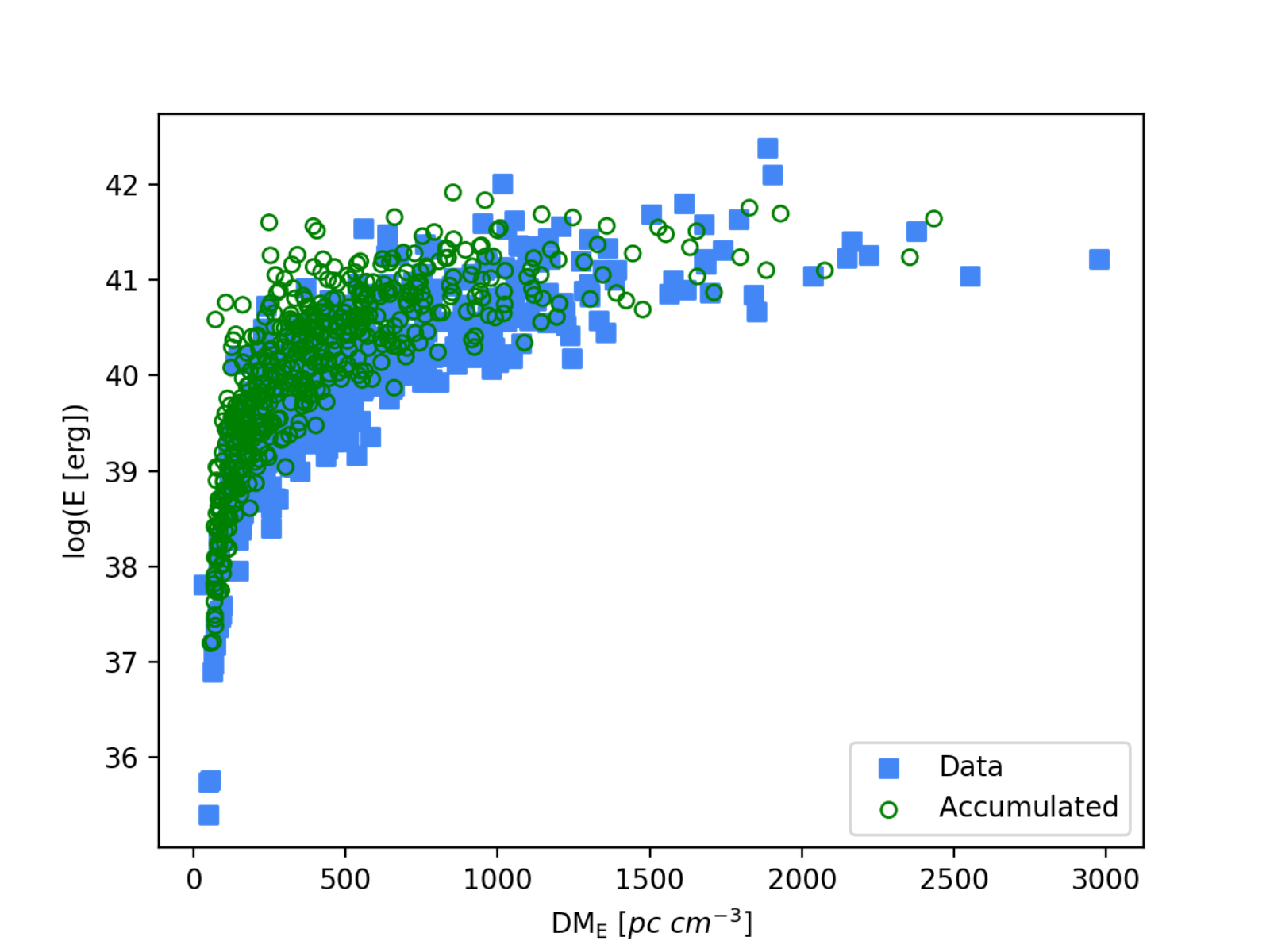}}
    \caption{Similar to Figure \ref{fig:SFH}, but for a test of an accumulated star formation model. All the simulations are scaled to the data. For the energy distribution model, we adopt $\alpha = 1.9$ and $\log E_c = 41$.}
\label{fig:accumulated}
\end{figure*}

Next, we test a $z$-distribution model that tracks the accumulated stars throughout history (green curves in Fig.\ref{fig:models}). \cite{hashimoto20} advocated a model with nearly constant FRB rate during the past 10 Gyr of look-back time. They interpreted this as FRBs tracking the accumulated stellar population. We therefore test such an accumulated SFH model in detail. As shown in Fig.\ref{fig:accumulated}, this model describes the data better than the SFH model. However, given a fluence distribution that is not rejected by the data, the resulting energy and $\rm DM_E$ distributions are still both rejected by the K-S test when tested against the data. Thus, we still reject this model to describe the CHIME FRB population. Nonetheless, since this model is a better representation of the data, it suggests that a large fraction of FRBs likely are produced by a stellar population that is significantly delayed with respect to star formation.

\subsection{Delayed models}
\label{section:delay}

We next explore a family of models that invoke a significant delay from star formation. As described in detail in \cite{zhangrc21}, we first generate $dN/(dt dV)$ based on the SFH model and calculate the look-back time $t_L$ distribution of the sample. Next, we introduce a distribution model for the delay time $\tau$, e.g. in the form of power-law, Gaussian, or lognormal functions \citep[e.g.][]{virgili11,sun15,wanderman15}. We then subtract the look-back time of the SFH model by the delay time $\tau$ for each mock FRB and finally obtain the look-back time distribution of the delayed population. The cases of negative $t_L$ are dropped out since they stand for future events. Finally, we convert the new $t_L$ distribution to the $dN/(dt dV)$ distribution of the new model, and simulate their redshift distribution $dN/(dt_{\rm obs} dz)$ using Eq.(\ref{eq:dN/dtdz}). The same approach is then applied to test the three criteria in $\log {\cal F}_\nu$, $\log E$, and $\rm DM_E$ distributions.

\begin{figure*}
    \centering
    \subfigure[]{\includegraphics[width=0.39\paperwidth]{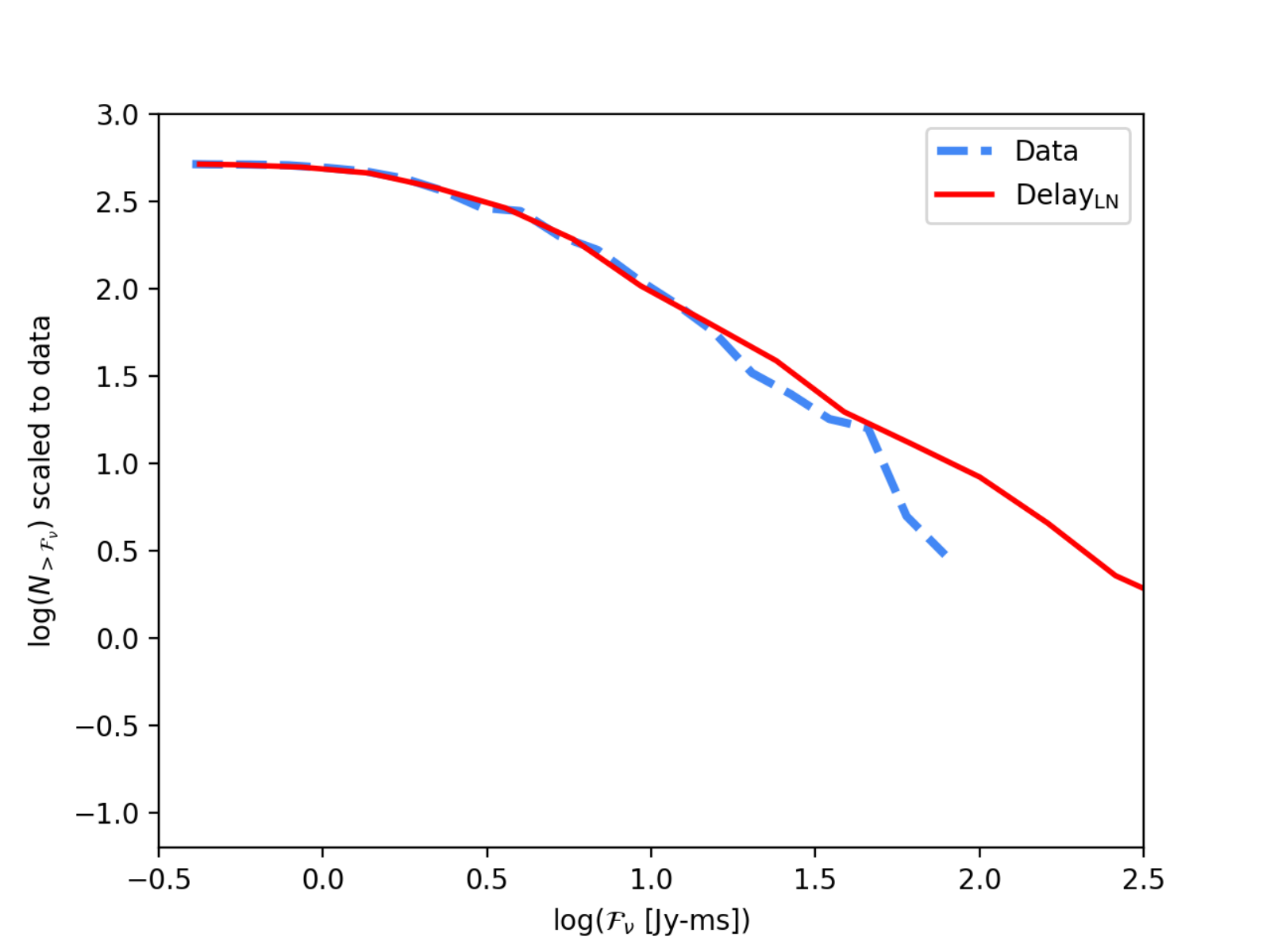}} 
    \subfigure[]{\includegraphics[width=0.39\paperwidth]{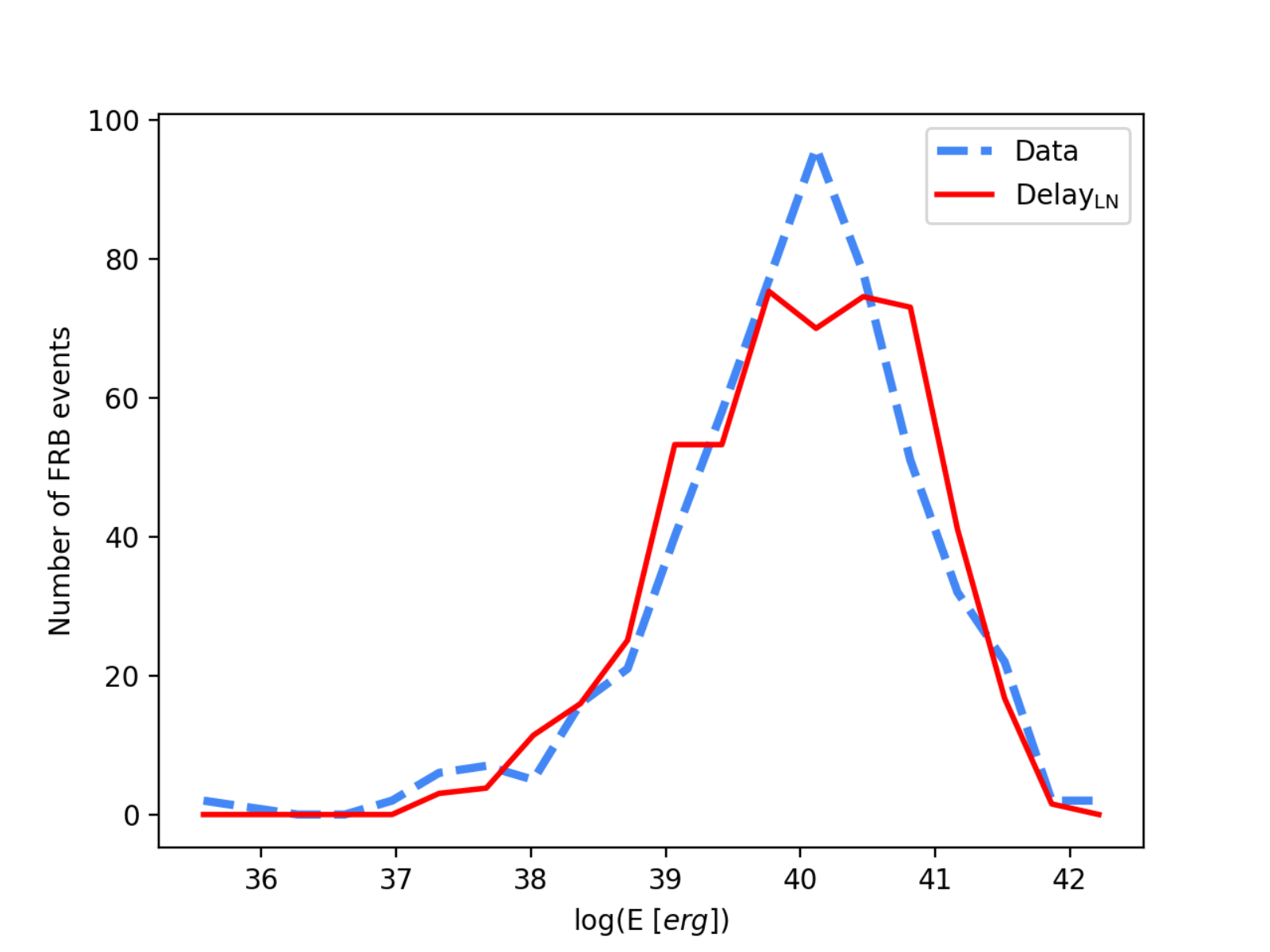}} 
    \newline
    \subfigure[]{\includegraphics[width=0.39\paperwidth]{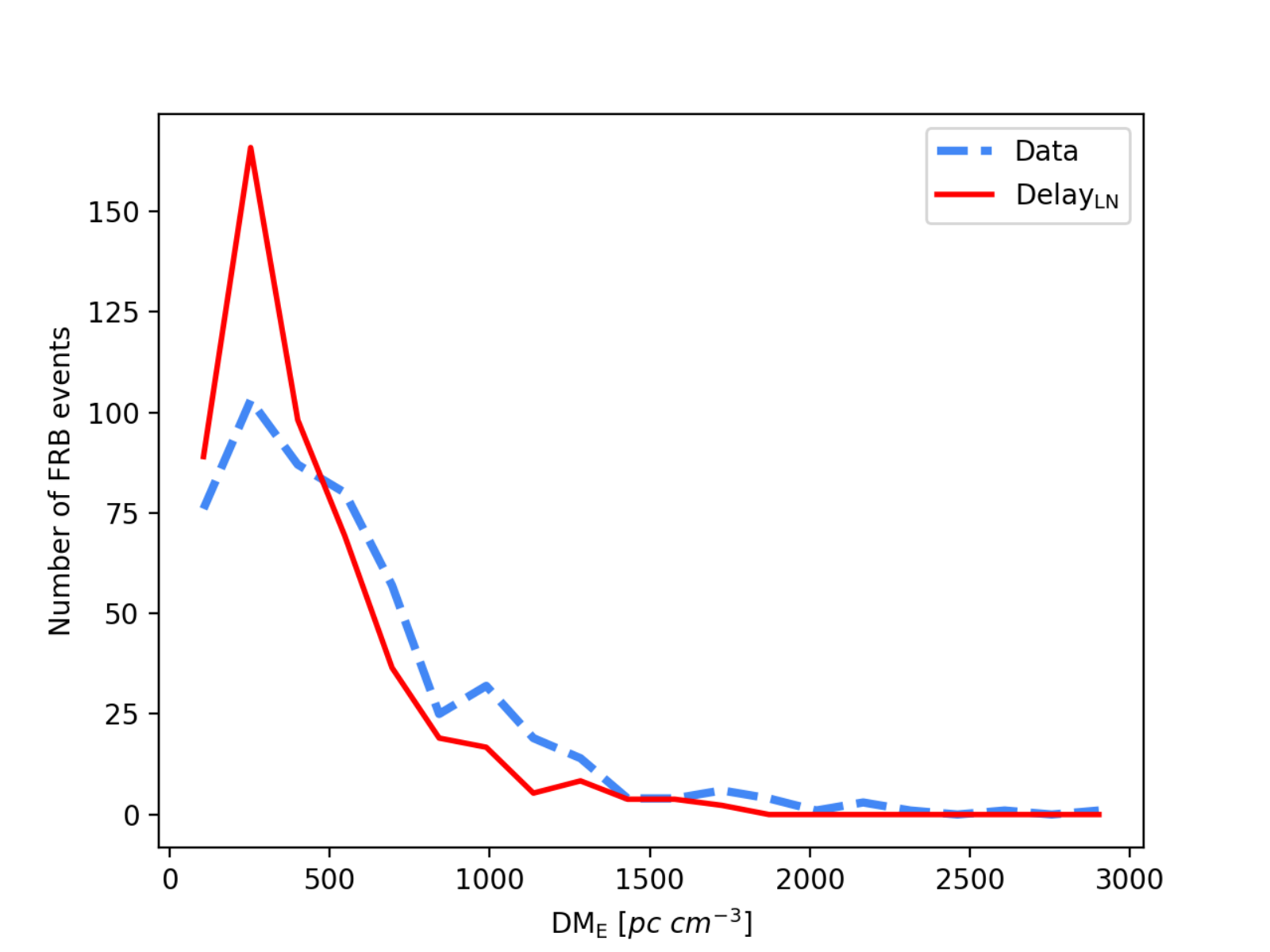}}
    \subfigure[]{\includegraphics[width=0.39\paperwidth]{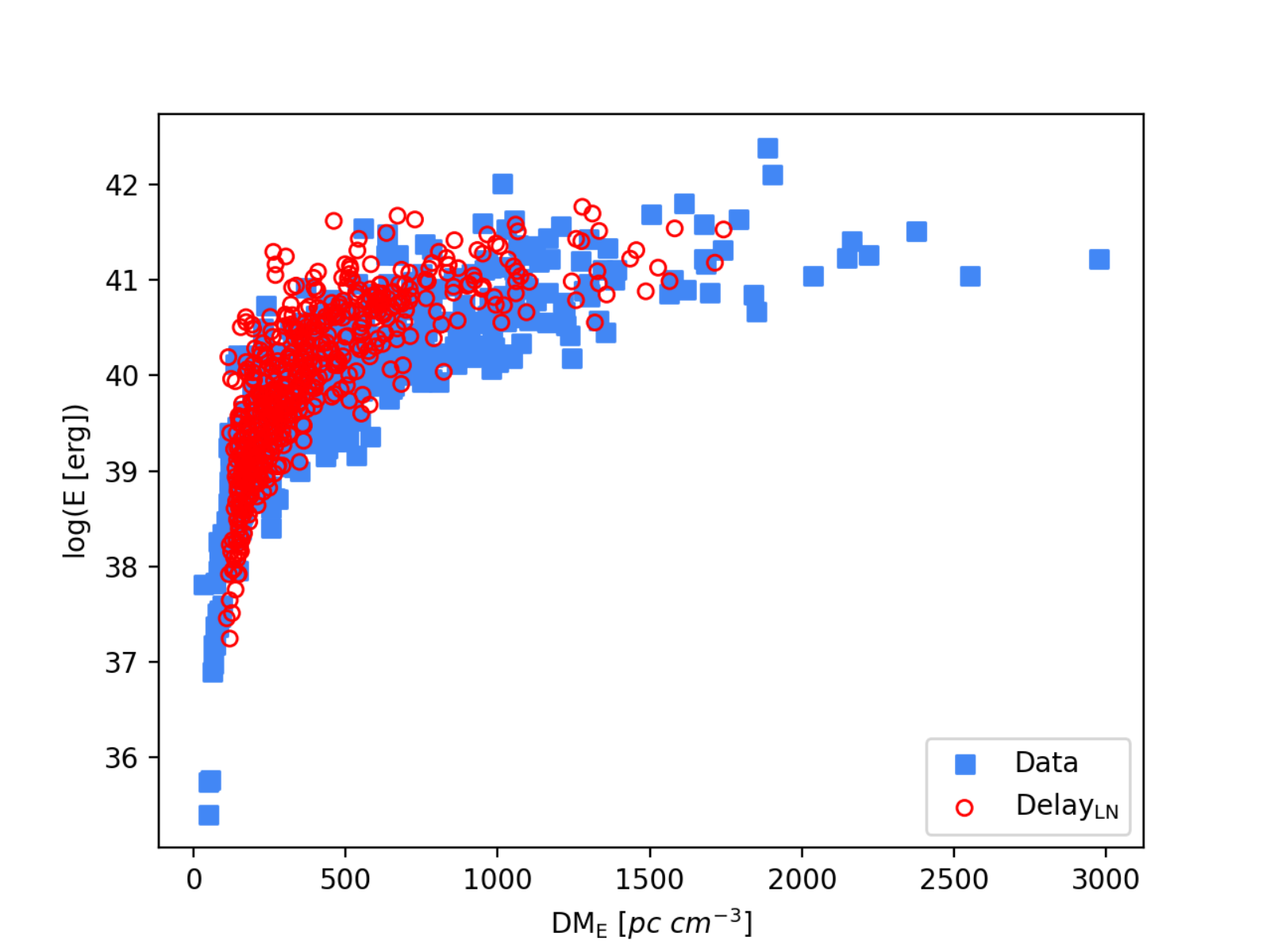}}
    \caption{Similar to Figure \ref{fig:SFH}, but for a test of a lognormal delay model with a central value of 10 Gyr and a standard deviation of 0.8 dex. All the simulations are scaled to the data. For the energy distribution model, we adopt $\alpha = 1.9$ and $\log E_c = 41$.}
    
\label{fig:merger}
\end{figure*}

Even though a family of delayed models was not rejected by the K-S test with the Parkes and ASKAP data as shown in \cite{zhangrc21}, we find that all of the models previously tested (with a characteristic delay timescale of $\sim (2-3)$ Gyr and consistency with the short GRB data) are rejected by the K-S test with the CHIME data sample. Rather, the data require a model with a much longer delay. Figure \ref{fig:merger} (also red curves in Fig.\ref{fig:models}) presents an example of such a lognormal delay model with a central value of 10 Gyr and a standard deviation of 0.8 dex. It reproduces the data much better than the SFH model, with both the fluence and energy criteria being not rejected by the K-S test. Even though the $\rm DM_E$ criterion is still rejected by the K-S test, the K-S statistic is much closer to the critical value (to not reject the null hypothesis) than the SFH model. We note that there is a wide range of the inferred $\rm DM_{host}$ in the FRB data, which would cause more complicated features in the modeled $\rm DM_E$ distribution than our simple model. This could partially account for the $\rm DM_E$ discrepancy between our model and the data. Since there are many parameters in the model and since our goal is to test the general trend of the models rather than identify model parameters, we do not make efforts to search for the best parameter set to satisfy the observational constraints. In any case, we can conclude that such a significantly delayed model offers a better description to the CHIME data than the SFH model.

\subsection{Hybrid model}

\begin{figure*}
    \centering
    \subfigure[]{\includegraphics[width=0.39\paperwidth]{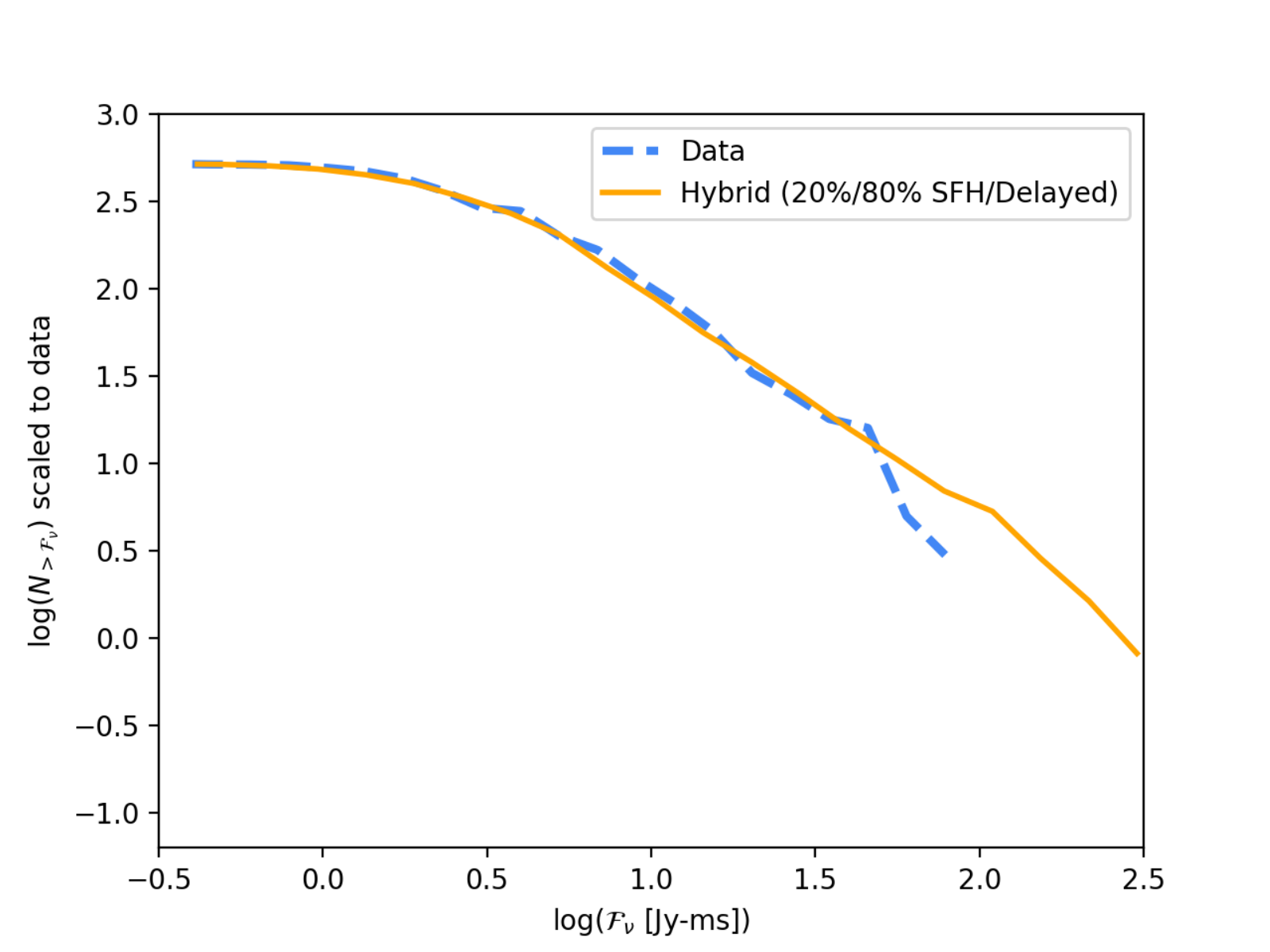}} 
    \subfigure[]{\includegraphics[width=0.39\paperwidth]{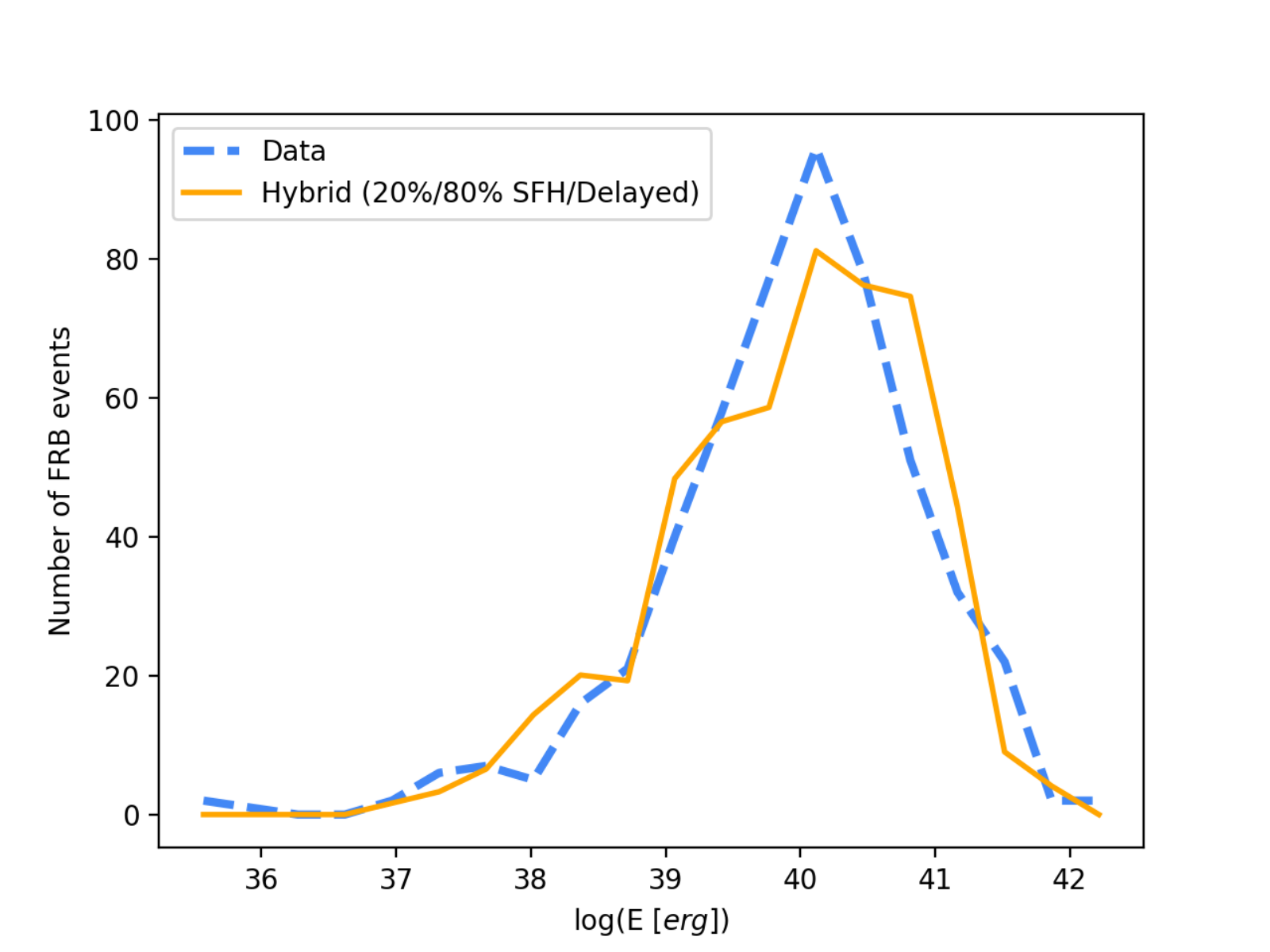}} 
    \newline
    \subfigure[]{\includegraphics[width=0.39\paperwidth]{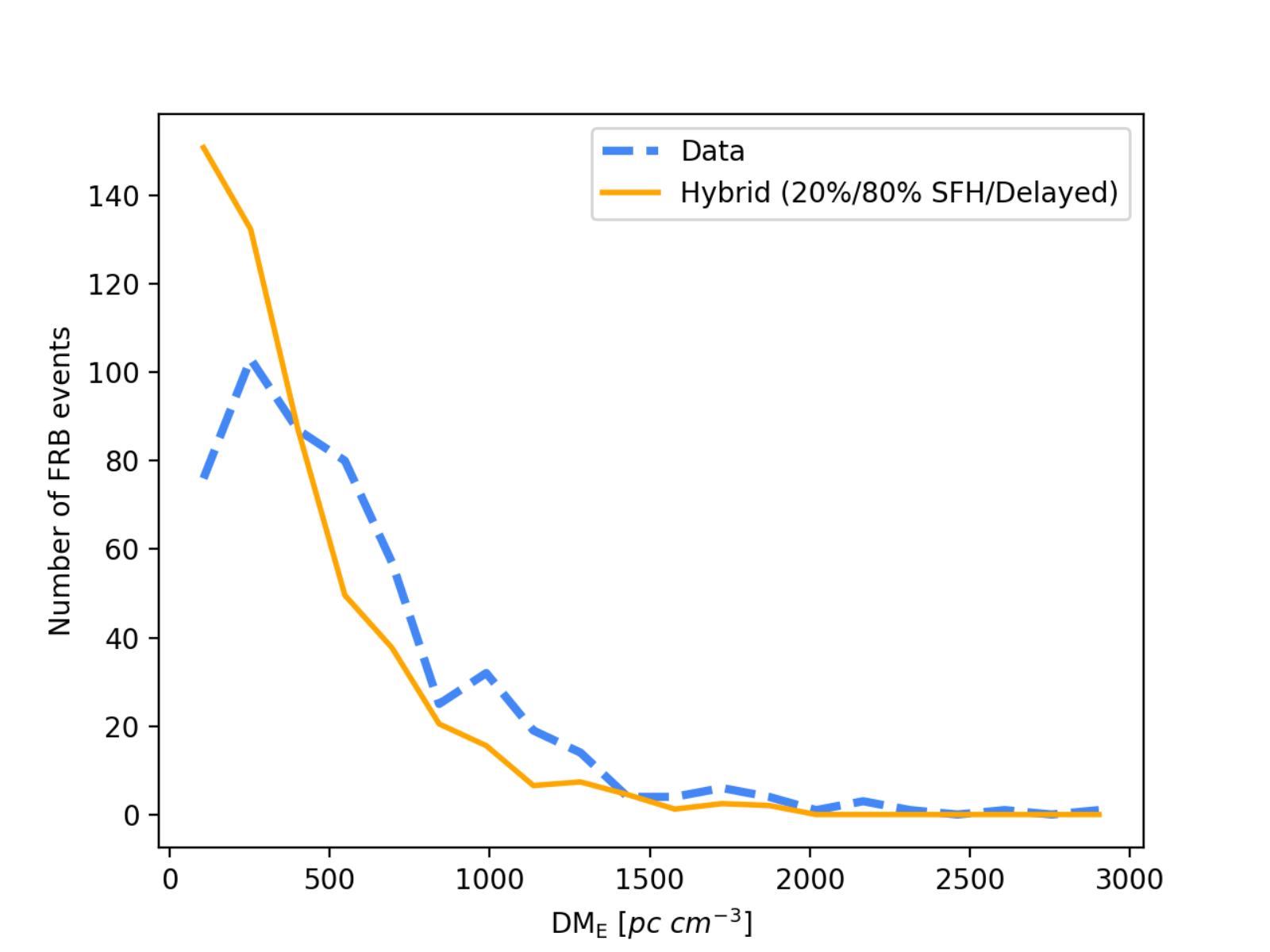}}
    \subfigure[]{\includegraphics[width=0.39\paperwidth]{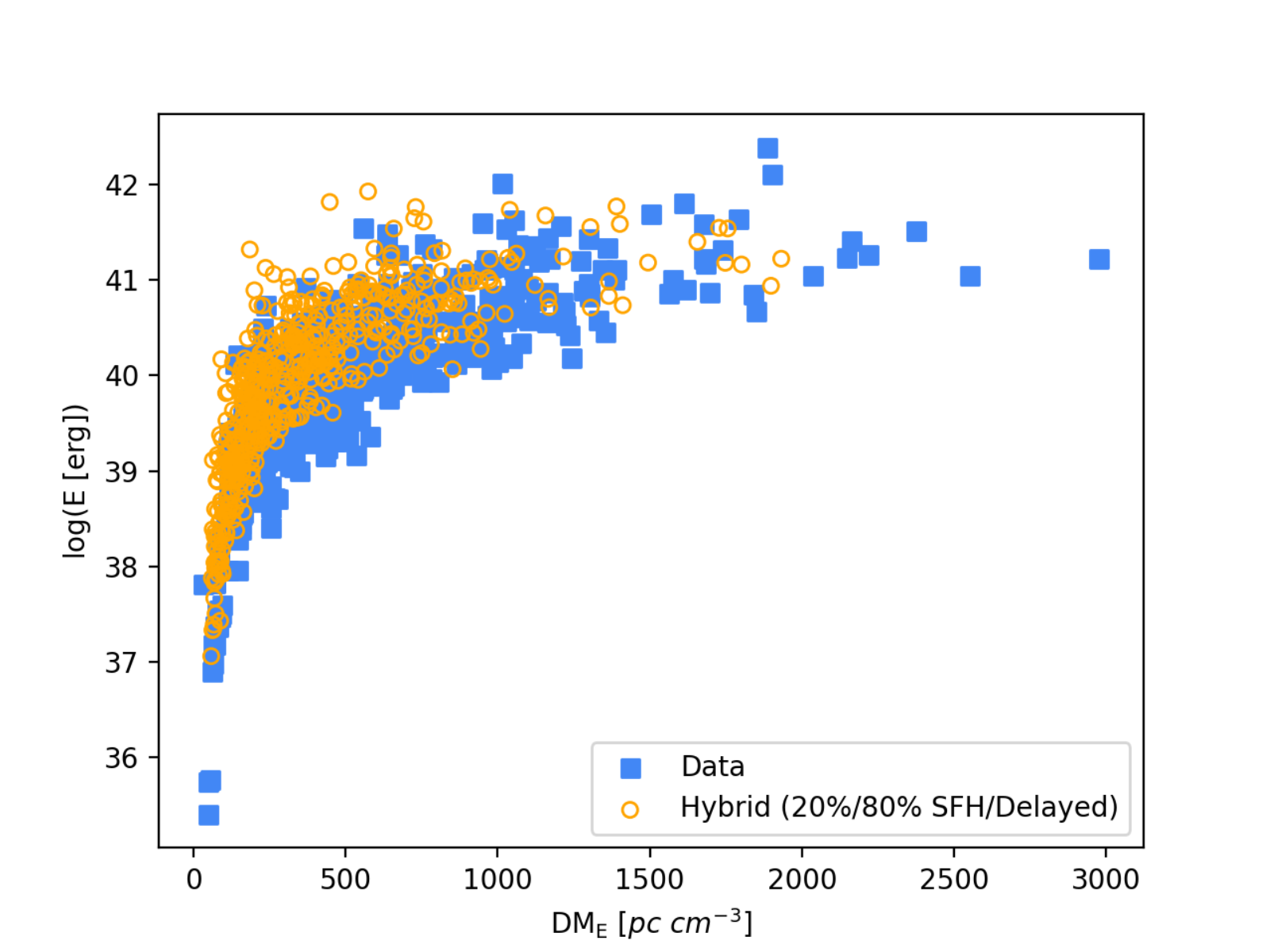}}
    \caption{Similar to Figure \ref{fig:SFH}, but for a test of a hybrid model that includes a young component tracking the SFH ($20\%$ of the total population) and an old component ($80\%$ of the total population) with lognormal delay with a central value of 13 Gyr and standard deviation of 0.8 dex. All the simulations are scaled to the data. For the energy distribution model, we adopt $\alpha = 1.9$ and $\log E_c = 41$.}
\label{fig:hybrid}
\end{figure*}

Finally, we test a family of models that includes the mixture of a young component that tracks star formation history and an old component that has a significant delay with respect to star formation (orange curves in Fig.\ref{fig:models}). Such a model is motivated by the discoveries of the Galactic FRB 200428 \citep{CHIME-SGR,STARE2-SGR}, which tracks star formation, and the M81 globular cluster FRB 20200120E \citep{bhardwaj21,kirsten21,nimmo21}, which tracks an old population. Since the SFH model fails badly, the proportion of the young population cannot be high. To compensate the high-DM events predicted from the young population, the old population in the hybrid model needs to have an even longer delay from star formation. We test a model with a lognormal delay distribution (central value = 13 Gyr, standard deviation = 0.8 dex) of the old population that is mixed with the SFH model of the young population. The proportions are 80\% for the delayed component and 20\% for the SFH component. The results are shown in Figure \ref{fig:hybrid}. We see similar results to that of the delayed model, with the fluence and energy criteria being not rejected, but $\rm DM_E$ still rejected.  It is clear that this model is also a much closer model to describe the data than the SFH model. Again, since there are even more parameters in this model than the delayed model and since our goal is to test the general trend of the models, we do not explore the best parameter set to fit the data. In any case, we deem that some hybrid models with a dominant delayed population component offer a better description to the CHIME data than the SFH model.

\section{Conclusions and discussion}\label{sec:conclusions}

Applying the Monte Carlo method developed in \cite{zhangrc21}, we have systematically tested a variety of FRB redshift distribution models against the first CHIME FRB catalog data. We draw the robust conclusion that the CHIME FRB population do not track the star formation history of the universe. The hypothesis that FRBs track all the stars formed in the universe (the accumulated SFH) model, despite describing the data better, is also rejected. Instead, the models that invoke a significant delay with respect to star formation seem to be working toward the correct direction in describing the data. A model that FRBs track a population of sources that have a significant delay ($\sim 10$ Gyr) with respect to star formation is one possibility. A hybrid model of young and old components is another possibility, but the old population needs to be the dominant component. Even though it is difficult to pin down the parameters due to many unconstrained parameters, this general trend is quite robust. Our conclusion is in general consistent with several previous suggestions that FRBs do not track star formation \citep[e.g.][]{cao17,hashimoto20, safarzadeh20}, even though an even longer characteristic delay time scale is needed to match the data better. Note that the reason that the delayed models are preferred is not because of their larger number of model parameters, but is rather because of the smaller peak value in the $\rm DM_E$ distribution (peaking around $250 \ {\rm pc \ cm^{-3}}$) of the CHIME sample compared with previous samples. 

Our results suggest that the ``most conservative'' scenario of FRB origin \citep{zhang20b}, i.e. magnetars can make them all, may have to be abandoned. Studies to interpret the M81 globular cluster FRB 20200120E \citep{kirsten21,kremer21,lu21} still invoke magnetars formed from other channels other than massive star deaths, e.g. binary white dwarf mergers, accretion induced collapses, and even binary neutron star mergers, to interpret FRB 20200120E. Since observationally, most known magnetars are formed from recent supernova explosions \citep{kaspi17}, one would expect that a small fraction of FRBs follow a delayed population if the magnetar hypothesis is correct. This is not what is inferred from the CHIME data, which requires that the delayed component is the dominant population. Challenges are raised to theorists regarding how to make FRBs with very old stars. One possibility is that FRBs are produced by reactivation of very old neutron stars across the universe, and old magnetars may fall into such a category \citep[e.g.][]{wadiasingh20,beniamini20b}. Other possibilities, e.g. FRBs being powered by interactions with old neutron stars \citep{mottez14,dai16,zhang17}, may also deserve reinvestigation.

\acknowledgements
We thank the referee for a constructive report and Wen-fai Fong, Jason Hessels, Clancy James, and Xavier Prochaska for helpful discussion and comments.



\end{document}